\documentclass[12pt,aps]{revtex4}
\usepackage{epsfig,amsmath,amssymb}

\newcommand{\eq}{{\, \equiv\, }}
\newcommand{\fr}[1]{%   arg1: var name
             \frac{#1}}
\newcommand{\be}{\begin{equation}}
\newcommand{\ee}{\end{equation}}
\newcommand{\bea}{\begin{eqnarray}}
\newcommand{\eea}{\end{eqnarray}}

\newcommand{\chibar}{\overline{\chi}}

\newcommand{\ket}{{\rangle}}
\newcommand{\e}{{\varepsilon}}
\newcommand{\bra}{{\langle}}
\newcommand{\gc}{\bra\fr{\alpha_s}{\pi}G^2\ket}
\newcommand{\qc}{\bra\,\overline{q}q\,\ket}
\newcommand{\bbar}{B\,-\,\overline{B}}
\newcommand{\ga}{{g_{{\cal A}}}}
\newcommand{\gat}{\tilde{g}_{{\cal A}}}
\newcommand{\ebar}{{\bar{\varepsilon}}}
\newcommand{\Qp}{Q^{(+)}_v}
\newcommand{\Qm}{Q^{(-)}_v}
\newcommand{\B}{P_{5i}^{(-)}\Sigma_{ii}^\dagger P_{5i}^{(+)}}
\newcommand{\Bs}{P_{i}^{(-)\mu}\Sigma_{ii}^\dagger P^{(+)}_{i\mu}}
\newcommand{\fm}{\left(\fr{f}{m}\right)}
\def\slash#1{#1 \hskip -0.5em / }

\begin{document}

\title{Non-factorizable effects in $\bbar$ mixing }
\author{A. Hiorth}
\email{aksel.hiorth@fys.uio.no}
\affiliation{Department of Physics, University of Oslo,
P.O.Box 1048 Blindern, N-0316 Oslo, Norway}
\author{J.O. Eeg}
\email{j.o.eeg@fys.uio.no}
\affiliation{Department of Physics, University of Oslo,
P.O.Box 1048 Blindern, N-0316 Oslo, Norway}
%\maketitle

\begin{abstract}

We study the $B$-parameter (``bag factor'') for $B -\overline{B}$ mixing
within a recently developed heavy-light chiral quark model.
 Non-factorizable contributions
 in terms of gluon condensates 
 and chiral corrections are calculated. 
In addition,
 we also consider $1/m_Q$ corrections within heavy quark effective 
field theory. Perturbative QCD effects below $\mu = m_b$ known from other
 work are also included. 
Considering two sets of input parameters, we find that the renormalization
 invariant $B$-parameter 
is  $\hat{B}=1.51 \pm 0.09$ for $B_d$ and
 $\hat{B}=1.40 \pm 0.16$  for  $B_s$.

\end{abstract}

\maketitle

\section{Introduction}

Studies of the neutral $K$-meson system have  played a major role in modern
 particle physics \cite{KKbar}.
Because of weak interactions,  a neutral $K$ meson may be transferred 
to a neutral  $\overline{K}$ meson. This process, known as
$K -\overline{K}$ mixing, determines
both the mass-difference  between the physical neutral
 states $K_L$ and $K_S$ and the dominating CP-violating effect in neutral
$K$-meson decays to pions (the $\varepsilon$-effect).
The neutral $B$-meson system has rather similar properties as the neutral 
$K$-system. The difference when going to $\bbar$ mixing is the   
importance of other KM quark mixing factors and other mass scales,
 in particular the 
$B$-mesons are about ten times heavier than the $K$-mesons.

In general, non-leptonic processes  may be described by an effective
Lagrangian which is a linear combination of quark operators. The 
(Wilson) coefficients of the operators  can be
 calculated in perturbation theory combined with the renormalization
 group equations \cite{weak}.
 At quark level, the leading order diagrams for $B -\overline{B}$ mixing
are given by the so called box diagram. This diagram has  double $W$- exchange
between two quark lines, and  generates an effective Lagrangian(Hamiltonian)
for the quark transition $\bar{b} d \rightarrow \bar{d} b$.
This Lagrangian has (for all practical purposes) only {\it one} operator
  times a
 Wilson coefficient containing  the effects of 
the virtual ($u,c,t$) quarks running in the loop. This Wilson coefficient
 has also been corrected for perturbative QCD effects within the
 renormalization group equations. Such calculations has been performed to
 next to leading order. For $B_s -\overline{B_s}$ mixing one considers
the corresponding $\bar{b} s \rightarrow \bar{s} b$ transition.

The difficult part is to
 calculate the matrix elements of the 
quark operators between the mesonic states, which is a non-perturbative issue.
 This has been done by lattice 
simulations \cite{latt,Mar-latt} or by  quark models \cite{Mel}.
 The hadronic matrix
 element is, as for $K -\overline{K}$ mixing, parameterized through the 
so called $B$- (``bag''-) parameter
which is by construction equal to one in the naive limit when vacuum 
states are inserted between
the quark currents in the $B -\overline{B}$ mixing operator.

In a previous paper \cite{BEFL}, $K -\overline{K}$ mixing was calculated
within a chiral quark model ($\chi$QM) combined with chiral perturbation 
theory. Within the $\chi$QM, non-factorizable contributions can also be 
calculated in terms of gluon condensates. The purpose of this paper is to
perform a similar analysis for $\bbar$ mixing. We are using a recently 
developed heavy-light chiral quark model (HL$\chi$QM) \cite{ahjoe}, 
 where non-factorizable
effects can be incorporated by means of gluon condensates and chiral loops.

\section{$\bbar$ mixing and heavy quark effective theory}

At quark level, the standard
effective Lagrangian describing $\bbar$ mixing is \cite{weak}:
\bea
{\cal L}_{eff}^{\Delta B= 2}&=&
 - \, \fr{G_F^2}{4\pi^2} M_W^2 \left(V_{tb}^*V_{tq}\right)^2 
\,S_0\left(x_t \right)\,\eta_B \,
b(\mu) \; Q(\Delta B= 2) \; , 
\label{LB2}
\eea
where  $G_F$ is Fermi's coupling constant, the $V$'s are 
KM factors \cite{KM}
 (for 
which $q=d$ or $s$ for $B_d$ and $B_s$ respectively)
 and $S_0$ is the Inami-Lim function \cite{inami} due to
 short distance electroweak
loop effects for the box diagram:
\begin{equation}
S_0(x)=\fr{4x- 11x^2+ x^3}{4(1- x)^2}-\fr{3x^3\,\text{Log}\,
x}{2(1- x)^3}\, .
\end{equation}
In our case, $x=x_t \equiv m_t^2/M_W^2$, where 
 $m_t$ is the 
%running 
top quark mass. 
%($m_t=165$ GeV in the $\overline{MS}$-scheme).
Because of its large mass, the top quark  gives the dominant
contribution. Also the $u$ and $c$ quarks are 
running in the loop, but these contributions are KM suppressed. The quantity 
$Q(\Delta B=2)$ is a four quark
operator :
\begin{equation}
Q(\Delta B = 2)= \overline{q_L}\,\gamma^\alpha\, b_L \;
\overline{q_L}\,\gamma_\alpha\, b_L \; ,
\label{QB2}
\end{equation} 
where $q_L$ $(b_L)$
 is the left-handed
projection of the $q$ $(b)$-quark field.
The quantities $\eta_B$ and $b(\mu)$ are calculated in perturbative
quantum chromodynamics (QCD). At the next to leading order (NLO)
analysis it is found that  
$\eta_B= 0.55\pm 0.01\,$ \cite{weak}.
Furthermore, for a renormalization point $\mu$ in perturbative QCD 
equal to or below $m_b$,
\bea
b(\mu) \, = \, \left[\alpha_s(\mu)\right]^{-6/23}
\left[1+\fr{\alpha_s(\mu)}{4\pi}J_5\right]\; ,
\label{Wils}
\eea
where $J_5$ = 1.63 in the naive dimension regularization scheme (NDR).
At $\mu~=~m_b$ $(=4.8~\text{GeV})$ one has $b(m_b)\simeq 1.56$.

The matrix element of the operator $Q(\Delta B = 2)$ between the meson
 states is parameterized 
by the bag parameter $B_{B_q}$ :
\begin{equation}
\bra B|Q(\Delta B= 2)|\overline{B}\ket \eq 
\fr{2}{3} f_B^2 M_B^2 B_{B_q}(\mu) \; \, .
\label{matrQ}
\end{equation}
By definition, $B_{B_q} =1 $ within  {\it naive factorization}, also named
vacuum saturation approach (VSA).
 This means  to insert a vacuum state between the two
 heavy-light currents in the operator $Q(\Delta B=2)$, and use
 the   matrix elements defining  the  decay 
constant $f_B$:
\bea
&&\bra 0|
\overline{q_L}\, \gamma^\mu\, b |\overline{B}(p)\ket=\fr{i}{2}\,f_B\,p^\mu\; 
\qquad\text{and}\qquad\bra B(p)|
\overline{q_L}\, \gamma^\mu\, b |0\ket=-\fr{i}{2}\,f_B\,p^\mu\; .
\label{fBdecay}
\eea
One may   combine naive factorization 
with the large $N_c$ expansion, where $N_c$ is the number of colours.
Then one  finds 
$B_{B_q} =  3(1+ 1/N_c)/4$, giving $B_{B_q}=3/4$ in the (naive) large
$N_c$ limit. We will see later that there are important non-factorizable
 contributions of order $1/N_c$.
In general, the matrix elements of  the operator $Q(\Delta B = 2)$
are dependent on the renormalization scale $\mu$, and thereby $B_{B_q}$
depends on $\mu$. 
 As for $K-\overline{K}$
mixing, one defines a renormalization scale independent quantity
\begin{equation}
\hat{B}_{B_q} \equiv b(\mu) B_{B_q}(\mu) \; .
\label{Bhat}
\end{equation}
Within lattice gauge theory,    values for  $\hat{B}_{B_q}$
 between 1.3 and 1.5 are obtained \cite{latt,Mar-latt}.

The mass difference between the weak eigenstates ($B_H$ and $B_L$) are
related to the bag parameter in the following way for $B_q = B_d, B_s$:
\begin{equation}
\Delta m_q=\fr{G_F^2}{6\pi^2}m_{B_q}f_{B_q}^2 \hat{B}_{B_q}\eta_B M_W^2\,S_0
\left(m_t^2/M_W^2\right)\,|V_{tq}^*V_{tb}|^2 \; .
\end{equation}
In order to extract the KM matrix elements it is crucial to have a
precise knowledge of the bag parameter $\hat{B}_{B_q}$, and the weak decay
constant $f_{B_q}$. 

The $b$ -quark is heavy compared to the typical hadronic scale of order 1 GeV,
where confinement and chiral symmetry breaking effects are essential.
 Perturbative effects below the $b$-quark
 scale may then be calculated down to 1 GeV by means of
 heavy quark effective theory (HQEFT. See \cite{neu} for
a review). Thus HQEFT also allows us to  evolve  the matrix element
(\ref{QB2}) from $\mu=m_b$ down to 1 GeV.

 HQEFT is a systematic expansion in
$1/m_b$. The heavy quark field $b(x)$ is replaced by a ``reduced''
field, $Q_v^{(+)}(x)$ or $Q_v^{(-)}(x)$, which is related to the full
 field the in following way:
\begin{equation}
Q_v^{(\pm)}(x)=P_{\pm}e^{\mp im_b v \cdot x}b(x) \, ,
\end{equation}
where $P_\pm$ are projecting operators $P_\pm=(1 \pm \gamma \cdot v)/2$. The
reduced field $Q_v^{(+)}$ can only annihilate heavy quarks. 
In order to describe 
heavy anti-quarks one has to use  
$Q_v^{(-)}$. In other words, $Q^{(+)}_v (Q_v^{(-)})$ annihilates (creates)
a heavy quark (anti-quark) with velocity $v$. The Lagrangian for heavy
quarks is ($Q_v = Q_v^{(\pm)}$):
\begin{equation}
{\cal L}_{\text{HQEFT}} =   \pm \overline{Q_{v}} \, i v \cdot D \, Q_{v} 
 + \frac{1}{2 m_Q}\overline{Q_{v}} \, 
\left( - C_M \frac{g_s}{2}\sigma \cdot G
 \, +   \, (i D_\perp)_{\text{eff}}^2  \right) \, Q_{v}
 + {\cal O}(m_Q^{- 2}) \; \, ,
\label{LHQEFT}
\end{equation}
where $D_\mu$ is the covariant derivative containing the gluon field
(eventually also the photon field), and 
$\sigma \cdot G = \sigma^{\mu \nu} G^a_{\mu \nu} t^a$, where 
$\sigma^{\mu \nu}= i [\gamma^\mu, \gamma^\nu]/2$, $G^a_{\mu \nu}$
is the gluonic field tensors, and $t^a$ are the colour matrices. 
This chromo-magnetic term has a factor $C_M$ which is  one at tree level,
 but slightly modified by perturbative QCD effects below the scale $m_b$ . 
It has been calculated to NLO \cite{neub,grozin1}.
 Furthermore,  
$(i D_\perp)_{\text{eff}}^2 =
C_D (i D)^2 - C_K (i v \cdot D)^2 $. At tree level, $C_D = C_K = 1$.
Here, $C_D$ is not modified by perturbative QCD, while $C_K$ is different 
from one due to perturbative QCD corrections \cite{GriFa}.
In our case, $m_Q = m_b$ is the heavy quark mass. 

Running from $\mu = m_b$ down to $\mu = \Lambda_\chi=1\,$ GeV,
 there will appear
more operators. Some stem from the heavy quark expansion itself and some
are  generated by
perturbative QCD effects. The $\Delta B=2$ operator in equation
(\ref{QB2}) for  $\Lambda_\chi < \mu < m_b$ can be written
 \cite{gimenez1,mannel,gimenez2} :
\bea
&Q(\Delta B=2) =&\,  C_1(\mu) \; Q_1  +    C_2(\mu) \; Q_2
+\fr{1}{m_b}\left( \sum_i^6 a_i(\mu) S_i(\mu)
\right.\nonumber\\&&\left.
+\sum_i^3\,h_i(\mu)X_i(\mu)\right)
\,+{\cal O}(1/m_b^2) \; \, .
\label{HQ}
\eea
The operator $Q_1$ is $Q(\Delta B=2)$ for $b$ replaced by $Q_v^{(\pm)}$,
while $Q_2$ is 
generated within perturbative QCD for $\mu < m_b$.  The operators
$S_i$ and $X_i$ are taking care of $1/m_b$ corrections. The quantities 
$C_1, C_2, a_i, h_i$ are Wilson coefficients. ($C_1=1 + {\cal O}(\alpha_s)$
and $C_2=0 + {\cal O}(\alpha_s))$. The explicit expressions for the
 operators are  
\bea
&Q_1  &=   2 \;  \overline{q_L}\,\gamma^\mu \,\Qp \; \; 
\overline{q_L}\,\gamma_\mu \,\Qm \, \; , 
\label{Q1}\\
&Q_2  &=   2 \; \overline{q_L}\,v^\mu \,\Qp \; \;
 \overline{q_L}\,v_\mu \,\Qm \; \, ,
\label{Q2}\\
&X_1&= 2 \;\overline{q_L}\,i\slash{D} \,\Qp \; \; 
\overline{q_L}\,\Qm\,+\, 2 \;\overline{q_L}\,iD^\mu \,\Qp \; \;
 \overline{q_L}\,\gamma_\mu\, \Qm
\,\nonumber\\
&&-2 \,i\,\varepsilon_{\lambda\mu\nu\rho}
\,v^\lambda\,\overline{q_L}\,iD^\mu\gamma^\nu \,\Qp \; \;
 \overline{q_L}\,\gamma^\rho\, \Qm\nonumber \\
&&+\,2 \;\overline{q_L}\,\Qp \; \; 
\overline{q_L}\,i\slash{D} \,\Qm\,+\, 2 \;\overline{q_L}\,\gamma_\mu\, \Qp \; \;
 \overline{q_L}\,iD^\mu 
\,\Qm\nonumber
\\&&
-i2 \,\,\varepsilon_{\lambda\mu\nu\rho}
\,v^\lambda\,\overline{q_L}\,\gamma^\nu\, \Qp \; \;
 \overline{q_L}\,iD^\mu\gamma^\rho \,\Qm   \; \, , \label{O1}\\
&X_2&=8\,\left[iv\cdot\partial(\overline{q_L}\,\Qp)\right]\overline{q_L}\,\Qm
+\,2\,\left[iv\cdot\partial(\overline{q_L}\gamma_\mu\,\Qp)\right]
\overline{q_L}\,\gamma^\mu\,\Qm\, ,\label{O2}\\
&X_3&=4\,\left[iv\cdot\partial(\overline{q_L}\gamma_\mu\,\Qp)\right]
\overline{q_L}\,\gamma^\mu\,\Qm \; \, . \label{O3}
\eea
The operators $S_i$ are nonlocal and is
a combination of the leading order operators $Q_{1,2}$ and a term of order
$1/m_Q$ from the effective Lagrangian (\ref{LHQEFT}):
\bea
\fr{S_1}{m_b}&=&i\int dy^4 
T\{Q_1(0),O_K(y)\} \; ,\nonumber \\ 
\fr{S_2}{m_b}&=&i\int d^4y 
T\{Q_2(0),O_K(y)\} \; ,\nonumber \\
\fr{S_3}{m_b}&=&i\int d^4y 
T\{Q_1(0),O_M(y)\} \; , \nonumber \\ 
\fr{S_4}{m_b}&=&i\int d^4y
T\{Q_2(0),O_M(y)\} \; ,\label{S}
\eea
where 
\bea
&O_K\,\eq&\,\fr{1}{2m_b}\left(\overline{\Qp}\,(iD_\perp)_{\text{eff}}^2\,\Qp
+\overline{\Qm}\,(iD_\perp)_{\text{eff}}^2\,\Qm\right) \; \, ,
\nonumber \\
&O_M\,\eq&\,-\fr{g_s}{4m_b}\left(\overline{\Qp}\,\sigma\cdot G\,
\Qp +\overline{\Qm}\,\sigma\cdot G\, \Qm\right) \; \,  ,
\eea
are the kinetic and magnetic operators of eq. (\ref{LHQEFT}).
There are no mixing between the local operators and the non-local
operators, since the local operators do not need the non-local ones
as counter-terms. The Wilson coefficients $a_i$ will then be the product
of $C_{1,2}$ and $C_{M,K}$.
The Wilson coefficients $C_1$ and $C_2$ have been calculated to NLO
 \cite{gimenez1,gimenez2} and for $\mu = \Lambda_\chi$,
$C_1(\Lambda_\chi)= 1.22$ and $C_2(\Lambda_\chi)= - 0.15$. 
The coefficients $h_{1,2,3}$ have been calculated to leading order (LO)
in \cite{mannel}, and the result at $\mu=\Lambda_\chi$ is $h_1=0.52$,
$h_2=-0.16$ and $h_3=-0.15$. 

\section{The heavy-light chiral quark model}

In order to calculate the matrix elements we will use the heavy-light
chiral quark model (HL$\chi$QM) recently developed in \cite{ahjoe}.
 This is a type of
  quark loop model \cite{chiqm,barhi,itCQM,effr} where
the quarks couples directly to the mesons at the scale of chiral
symmetry breaking
$\Lambda_\chi$, which we put equal to 1 GeV. What makes our model \cite{ahjoe}
distinct from other similar models is that it
 incorporates soft gluon effects in terms of the gluon condensate
with lowest dimension \cite{BEFL,pider,epb,BEF,EHP}.
 The term in the Lagrangian describing this
interaction can be obtained as a mean-field approximation of 
the (extended) Nambu-Jona-Lasinio model (NJL) \cite{bijnes,effr}.

In this section we will give a short presentation of the HL$\chi$QM.
 In the next section  we will use the model \cite{ahjoe} to calculate 
non-factorizable soft gluon effects in $\bbar$ mixing.

The Lagrangian for the HL$\chi$QM is 
\begin{equation}
{\cal L}_{\text{HL$\chi$QM}} =  {\cal L}_{\text{HQEFT}} +
  {\cal L}_{\chi\text{QM}}  +   {\cal L}_{\text{Int}} \; \, .
\label{totlag}
\end{equation}
The first term is given in equation (\ref{LHQEFT}).
The light quark sector is described by the chiral quark model ($\chi$QM),
having a standard QCD term and a term describing interactions between
quarks and  (Goldstone) mesons: 
\begin{equation}
{\cal L}_{\chi\text{QM}} =  
\chibar \left[\gamma^\mu (i D_\mu   +    {\cal V}_{\mu}  +  
\gamma_5  {\cal A}_{\mu})    -    m \right]\chi 
  -     \chibar \, \widetilde{M_q} \, \chi \;  , 
\label{chqmR}
\end{equation}
where $\chi_{L,R}$ are the flavour rotated quark fields given by:
\begin{equation}
\chi_L  =   \xi^\dagger q_L \quad ; \qquad \chi_R  =   \xi q_R \quad ; \qquad 
\xi \cdot \xi  =   \Sigma \; .
\label{rot}
\end{equation}
where $q^T  =  (u,d,s)$ are the light quark fields. The left- and
 right-handed
 projections $q_L$ and $q_R$ are transforming after $SU(3)_L$ and $SU(3)_R$
respectively.
The quantity $\xi$ is a 3 by 3 matrix containing
 the (would be)  Goldstone octet ($\pi, K, \eta$) :
\begin{equation}\label{Pi}
\xi=e^{i\Pi/f}\, \quad
 \text{where}\, \quad \Pi=\fr{\lambda^a}{2}\phi^a(x) = 
\frac{1}{\sqrt{2}} \left[\begin{array}{ccc} \fr{\pi^0}{\sqrt{2}}+\fr{\eta_8}{\sqrt{6}} & \pi^+
&K^+\\ \pi^-&-\fr{\pi^0}{\sqrt{2}}+\fr{\eta_8}{\sqrt{6}} & K^0\\
K^- &\overline{K^0}& -\fr{2}{\sqrt{6}}\eta_8\end{array}\right] \; ,
\end{equation}
where $f$ is the bare pion decay constant.
In (\ref{chqmR}), $m$ is the ($SU(3)$ -  invariant) constituent quark 
mass for light quarks, and  $\widetilde{M_q}$  
contains the current quark mass matrix ${\cal M}_q$ and the field $\xi$:
\bea
&&\widetilde{M_q} \eq \widetilde{M}_q^V   +    \widetilde{M}_q^A \gamma_5  \;
 , \; \; \text{where} \label{cmass}\\
\widetilde{M}_q^V \, \eq \, 
\fr{1}{2}(\xi^\dagger {\cal M}_q^\dagger\xi^\dagger \;
 +&&\xi {\cal M}_q\xi )\quad\text{and}\quad 
\widetilde{M}_q^A\eq -\fr{1}{2}(\xi^\dagger {\cal M}_q^\dagger\xi^\dagger
 -\xi {\cal M}_q\xi) \; .
\label{masst}
\eea 
The vector and axial vector fields
${\cal V}_{\mu}$ and  
${\cal A}_\mu$ in (\ref{chqmR})  are given by:
\begin{equation}
{\cal V}_{\mu}\eq \fr{i}{2}(\xi^\dagger\partial_\mu\xi
+\xi\partial_\mu\xi^\dagger 
) \qquad ;  \qquad  
{\cal A}_\mu\eq  -  \fr{i}{2}
(\xi^\dagger\partial_\mu\xi
-\xi\partial_\mu\xi^\dagger) \; \, .
\label{defVA}
\end{equation}
Furthermore, the covariant derivative $D_\mu$ in (\ref{chqmR})
contains the soft gluon field forming the gluon condensates. The gluon 
condensate contributions are calculated by Feynman diagram techniques
as in \cite{BEFL,ahjoe,epb,BEF}. They may also be calculated
by means of heat kernel techniques as in \cite{pider,bijnes,ebert3}.

The interaction between heavy meson fields and heavy quarks are
described by the following Lagrangian :
\begin{equation}
{\cal L}_{Int}  =   
 -   G_H \, \left[ \chibar_a \, \overline{H_a^{(\pm)}} 
\, Q^{(\pm)}_{v} \,
  +     \overline{Q_{v}^{(\pm)}} \, H_a^{(\pm)} \, \chi_a \right]
\, + \,  \fr{1}{2G_3}Tr\left[ \overline{H_{v}^{a}}\,  H_{v}^{a}\right] \; 
 \, ,
\label{Int}
\end{equation}
where $G_H$ and $G_3$ are  coupling constants and
 $H_a^{(\pm)}$ is the heavy meson field  containing
 a spin zero and spin one boson:
\begin{eqnarray}
&H_a^{(\pm)} & \eq  P_{\pm} (P_{a \mu}^{(\pm)} \gamma^\mu -     
i P_{5 a}^{(\pm)} \gamma_5)\; \; \; , \; \; 
\overline{H_a^{(\pm)}}
\eq  \gamma^0 (H_a^{(\pm)})^\dagger \gamma^0
\; . \label{barH}
\end{eqnarray}
The fields $P^{(+)}(P^{(-)})$ annihilates (creates) a heavy meson containing
 a heavy quark (anti quark) with velocity  $v$. 

Integrating out the quarks by using (\ref{LHQEFT}), (\ref{chqmR}) and 
(\ref{Int}), the effective Lagrangian up
to ${\cal O}(m_Q^{-1})$ can be written as 
\cite{itchpt,ahjoe}:
\begin{equation}
{\cal L} =  \mp Tr\left[\overline{H^{(\pm)}_{a}}
\left(iv\cdot {\cal D}_{ba} - \Delta_Q\right)
H^{(\pm)}_{b}\right]\, -\, 
\ga \, Tr\left[\overline{H^{(\pm)}_{a}}H^{(\pm)}_{b}
\gamma_\mu\gamma_5 {\cal A}^\mu_{ba}\right]\, 
,\label{LS1}
\end{equation}
where $i{\cal D}^\mu_{ba}=i \delta_{ba} D^\mu-{\cal V}^\mu_{ba}$.
The term proportional to the quark-meson mass difference 
$\Delta_Q =M_H-m_Q$ in (\ref{LS1}) is irrelevant for us
due to the reparametrization invariance \cite{neu}. Also, it does not
enter our loop integrals  because our heavy meson
fields are attached to our quark loops at zero external momentum. 
(The external momentum includes the piece $v^\mu \Delta_Q$).
As shown in \cite{ahjoe}, the term $\sim 1/G_3$ in (\ref{Int}) is
related to $\Delta_Q$, and this term is also irrelevant within
 the present paper.

To obtain (\ref{LS1}) from the HL$\chi$QM  one encounters divergent loop
integrals, which might be quadratic-, linear- and
logarithmic divergent.
For the kinetic term in (\ref{LS1}) we obtain the identification:
\begin{equation}
 -  iG_H^2N_c \, \left(I_{3/2}  +   2mI_2  
-i  \fr{(3 \pi -8)}{384 m^3 N_c}\gc \right) =   1 \; \, ,
\label{norm}
\end{equation}
where $I_{3/2}$ and $I_2$ are the linear and logarithmic divergent
integrals respectively, and $\gc$ is the gluon condensate.
To obtain  the axial vector term proportional to $g_{\cal A}$, we obtain
a similar condition, and combining it with (\ref{norm}), we obtain 
 for  the axial vector term 
\begin{equation}
\ga=1+\fr{4}{3}iG_H^2N_c \left(I_{3/2}-\fr{im}{16\pi}\right) \; \, ,
\label{ga}
\end{equation}
such that the (formally) linear divergent integral $I_{3/2}$ is related to 
the strong axial coupling $\ga$ (or strictly speaking, 
its deviation from one).
Analogously,
within the pure light quark sector (the $\chi$QM), it is well known
that the quadratic and
logarithmic divergent integrals are related to the quark condensate and
the bare decay constant
$f$, respectively \cite{chiqm,pider,epb,BEF,ebert3}:
\bea
&\qc &= -4imN_cI_1-\fr{1}{12m}\gc \; \, ,\label{I1}\\
&f^2 &=    -  i4m^2N_cI_2 +  \fr{1}{24m^2}\gc \; \, .
\label{I2}
\eea
The divergent integrals $I_1,\, I_2$ and $I_{3/2}$ are listed in
appendix \ref{app:loop}. The effective coupling $G_H$ describing the
interaction between the quarks and heavy mesons can be expressed in terms
of $m$, $f$, $g_{\cal A}$, and the  mass splitting between
 the $1^-$ state and $0^-$ state.
Using (\ref{norm}), (\ref{ga}), (\ref{I2})
 one finds a  relation between this mass-splitting and the
gluon condensate   via the chromomagnetic 
interaction in (\ref{LHQEFT}) \cite{ahjoe} :
\begin{equation}
\label{gcgh}
\gc  =   \fr{16 f^2}{\pi \eta} \, \fr{\mu_G^2}{\rho}  \; \, , \qquad  
G_H^2  =   \fr{2m}{f^2} \, \rho \; \, ,
\qquad  \eta \, \eq\, 
\fr{(\pi +  2)}{\pi}C_M(\Lambda_\chi) \; \, ,
\end{equation}
where 
\begin{equation}
\label{rho}
\rho \, \eq \, \fr{(1 +  3g_{\cal A}) + 
 \frac{\mu_G^2}{\eta \, m^2}}{4 (1 + \frac{N_c m^2}{8 \pi f^2} )}
\; \; \; , \; \; \; \mu_G^2(H) \, =  \, \frac{3}{2} m_Q (M_{H^*} -  M_H) .  
\end{equation}
In the limit where only the leading logarithmic integral $I_2$ is kept
we obtain :  
\begin{equation}
g_{\cal A} \rightarrow \, 1 \; , \qquad
\rho \rightarrow \, 1 \; , \qquad  
G_H \; \rightarrow \; G_H^{(0)} \, \eq \, \fr{\sqrt{2m}}{f} \; \, . 
\label{IRlim}
\end{equation}
Note that $g_{\cal A}= 1$ is the non-relativistic value \cite{itchpt}.
We observe that the mass-splitting between $H$ and $H^*$
sets the scale of the gluon condensate.
This means that, while in \cite{BEF} the gluon condensate was fitted to the
$K \rightarrow (2 \pi)_{I=2}$ amplitude, it is here determined in the  
strong sector alone (with a slightly lower value than in \cite{BEF}).
 
The $1/m_Q$ corrections to the strong Lagrangian have been calculated in
\cite{ahjoe}. They may  formally be put 
into spin dependent renormalization factors.
This means  that (\ref{LS1}) is still valid with the replacement
$H^{\text{r}}= H \, (Z_H)^{-\frac{1}{2}}$, where $Z_H$
 and the renormalized (effective) coupling $\gat$ are defined as:
\bea
Z_H^{-1}  &=& 1+\fr{\varepsilon_1-2d_M\varepsilon_2}{m_Q} \; \;  , \\
\gat &=& \ga \left(1-\fr{1}{m_Q}(\varepsilon_1-2d_{\cal A}\varepsilon_2)\right)
-\fr{1}{m_Q}(g_1-d_{\cal A}g_2) \; \, ,
\eea
where
\begin{equation}
d_M=\begin{cases}\,\,\,\,\, 3\quad\text{for}\quad 0^-\\
                 -1\quad\text{for}\quad 1^-
\end{cases}\quad\quad
d_{\cal A}=\begin{cases}\,\,\,\, 1\quad\text{for}\quad H^*H\,\,\,\quad\text{coupling}\\
              -1\quad\text{for}\quad H^*H^*\quad\text{coupling}
\end{cases}
\end{equation}
and :
\bea
\varepsilon_1&=&-m+G_H^2\left(\fr{\qc}{4m}+f^2+\fr{N_cm^2}{16\pi}+\nonumber
\fr{C_K}{16}(\fr{\qc}{m}-f^2)\right.\\&&\left.\qquad\qquad\quad
+\fr{1}{128 m^2}(C_K+8-3\pi) \gc\right) \; , \\
g_1&=&\,m-G_H^2\left(\fr{\qc}{12m}+\fr{f^2}{6}+
\fr{N_cm^2(3\pi+4)}{48\pi}- \nonumber
\fr{C_K}{16}(\fr{\qc}{m}+3f^2)\right.\\&&\left.\qquad\qquad\quad
+\fr{1}{64 m^2}(C_K-2\pi)\gc\right) \; , \\
g_2&=& \fr{(\pi + 4)}{(\pi+2)} \fr{\mu_G^2}{6 m}
\; .
\eea

\section{Bosonizing $Q(\Delta B=2)$}

In this section we will discard $1/m_Q$ terms. We are then left with
the operators $Q_{1,2}$ defined in equation (\ref{Q1}) and
(\ref{Q2}).
In order to find the matrix element of $Q_{1,2} \,$, one uses the 
following relation between the generators of $SU(3)_c$ ($i,j,l,n$
are colour indices running from 1 to 3):
\begin{equation}
\delta_{i j}\delta_{l n}  =   \fr{1}{N_c} \delta_{i n} \delta_{l j}
 \; +  \; 2 \; t_{i n}^a \; t_{l j}^a \; ,
\label{fierz}
\end{equation}
where $a$ is an index running over the eight gluon charges. This
 means that  by means of a Fierz transformation, the operator $Q_1$
 in (\ref{Q1}) may  also be written in the following way : 
\begin{equation}
Q_1  =   \fr{2}{N_c}\, \overline{q_L} \,\gamma^\mu\, \Qp\,
\overline{q_L}\, \gamma_\mu\, \Qm
\,+\,  4\, \overline{q_L}\, t^a\, \gamma^\mu  \,\Qp \,
 \overline{q_L} \,t^a\, \gamma_\mu\, \Qm \, ,
\label{Q1Fierz}
\end{equation}
and similarly for $Q_2$. 

\begin{figure}[t]
\begin{center}
   \epsfig{file= 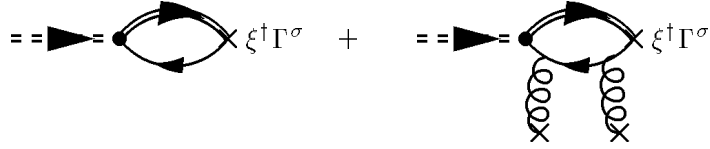,width=10cm}
\caption{Diagrams for bosonization of the left handed quark current}
\label{fig:f+}
\end{center}
\end{figure}

The first (naive) step to calculate the matrix element of a four quark 
operator like $Q_1$ is by inserting vacuum states between the two currents.
This vacuum insertion approach (VSA)
corresponds to bosonizing the two currents in $Q_1$ and multiply them, as 
mentioned below eq. (\ref{matrQ}).
For one current, visualized in figure  \ref{fig:f+},  
 one obtains \cite{itchpt,ahjoe} :
\begin{equation}
 \overline{q_L} \,\gamma^\mu\, Q_v^{(\pm)} \;  \longrightarrow \;
    \fr{\alpha_H}{2} Tr\left[\xi^{\dagger}_{hf}\gamma^\alpha
L \,  H_{h}^{(\pm)} \right]
  \; ,\label{J(0)}
\end{equation}
Using the relations (\ref{norm}) - (\ref{I2}) for the divergent integrals,
and also eq. (\ref{gcgh}), we obtain \cite{ahjoe}:
\begin{equation}\label{qcrel}
\alpha_H=\fr{G_H}{2}\left(-\fr{\qc}{m}-2f^2(1-\fr{1}{\rho})+
\fr{(\pi-2)}{16m^2}\gc\right) \; .
\end{equation}

This bosonization has to be compared with the 
 matrix elements defining  the meson decay 
constant $f_B$ given in eq. (\ref{fBdecay}). In those relations,
$b$ is the full quark field. Within HQEFT this matrix element 
will,  below the renormalization scale
 $\mu  =   m_Q \, (=m_b)$, be modified in the following way:
\bea
&&\bra 0|
\overline{q_L}\, \Gamma^\mu\, \Qp |\overline{B}(p)\ket=
\fr{i}{2}\,f_B\,M_B\,v^\mu \; 
\quad\text{and}\quad\bra B(p)|
\overline{q_L}\, \Gamma^\mu\, \Qm |0\ket=-\fr{i}{2}\,f_B\,M_B\,v^\mu \; , 
\eea
where \cite{neu}
\begin{eqnarray}
\Gamma^\mu \,\eq\, C_\gamma (\mu )\,\gamma^\mu\,+\,  
C_v(\mu )\, v^\mu\; .
\label{Gamma}
\end{eqnarray}
The coefficients $C_{\gamma,v}(\mu)$ are determined 
by QCD renormalization for  $\mu < m_Q$. They have been calculated to
NLO and the result is the same in $MS$ and $\overline{MS}$
scheme \cite{Cgamma}. 
In HL$\chi$QM the decay constant $f_B$ can be calculated and the
result is \cite{ahjoe} :
\begin{equation}
\alpha_H =  \fr{f_B\sqrt{M_B}}{C_\gamma (\mu ) + C_v(\mu )} =
\fr{f_{B^*}\sqrt{M_{B^*}}}{C_\gamma(\mu)} \;\;  .
\label{fb}
\end{equation}

\begin{figure}[t]
\begin{center}
   \epsfig{file=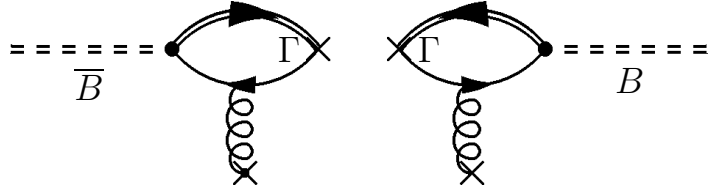,width=10cm}
\caption{Nonfactorizable contribution, $\Gamma\eq t^a \,\gamma^\mu\,L$}
\label{fig:bbargg}
\end{center}
\end{figure}

 The second matrix element in (\ref{Q1Fierz}) is genuinely non-factorizable,
 and we have to go beyond the VSA.
However, in the approximation where only the lowest gluon condensate is
 taken into account, the last term in (\ref{Q1Fierz}) can be written in a 
{\it quasi-factorizable} way by
bosonizating the heavy-light coloured current 
with an extra  colour matrix $t^a$ inserted and with an extra gluon 
emitted 
as shown in figure \ref{fig:bbargg}. Calculation of this diagram is
straightforward when using the light quark propagator with just one
 soft gluon emitted :
\begin{equation}
S_G(k) \eq   \fr{g_s}{4} G^b_{\alpha \beta} t^b
\left[ \sigma^{\alpha \beta} (\gamma \cdot k +  m)  +   
(\gamma \cdot k +  m) \sigma^{\alpha \beta}\right] (k^2- m^2)^{- 2} \; \; .
\label{S1G}
\end{equation}
 The part of the diagram to the left in figure
\ref{fig:bbargg} then gives  the bosonized coloured current:
\bea
\left(\overline{q_L} t^a\,\gamma^\alpha Q_v^{(\pm)}\right)_{1G} 
\;   \longrightarrow \;  
 -\fr{G_H g_s}{8}\,G_{\mu\nu}^a
\;Tr\left[\xi^\dagger \gamma^\alpha  L \, H^{(\pm)}
\left(\pm i\,I_2\left\{\sigma^{\mu\nu}, \gamma \cdot v\right\}+ \,
\fr{1}{8\pi} \sigma^{\mu\nu} \right)\right] \; ,
\label{1G}
\eea
where
$I_2$ is to be identified with $f^2$ by the use of equation (\ref{I2}).
The result for the right part of the diagram with $\bar{B}$ replaced by 
$B$  is obtained by just
changing the sign of $v$ and letting ${P_5^{(+)}}\rightarrow {P_5^{(-)}}$
(remembering that ${P_5^{(-)}}$ {\it creates} a meson with a heavy anti
quark).
 Multiplying the coloured currents,  we obtain for the non-factorizable
part of $Q_1$ and $Q_2$ to first order in the gluon condensate :
\bea
&&C_1\,\overline{q_L} t^a\,\gamma^\mu Q^{(+)}_v \;
\overline{q_L} t^a\,\gamma_\mu\,Q^{(-)}_v   
+C_2\,\overline{q_L} t^a\,v^\mu Q^{(+)}_v \;
\overline{q_L} t^a\,v_\mu\, Q^{(-)}_v  \nonumber \\
&& \rightarrow \; -\fr{\beta_B}{4}\gc
\left(C_1\B\,+(C_1-\fr{1}{3}C_2)\,\Bs\right) \; \, ,
\eea
where 
\begin{equation}
\beta_B\eq\fr{G_H^2}{128}\left\{1+\fr{4\pi}{N_c}\left(\fr{f}{m}\right)^2
+\fr{8\pi^2}{N_c^2}\left(\fr{f}{m}\right)^4\right\} \; \, ,
\label{BetaB}
\end{equation}
and  $\Sigma = \xi^2$, where $\xi$ is given in Eq. (\ref{Pi}).
Note there is no sum over $i$, $i=2,3$ for $q=d,s$ respectively.

\begin{figure}[t]
\begin{center}
   \epsfig{file=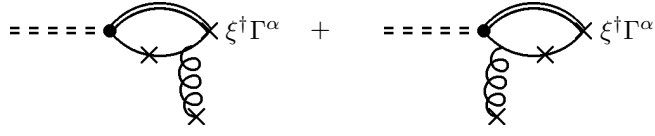}
\caption{Mass insertion in the nonfactorizable part of the current}
\label{fig:count}
\end{center}
\end{figure}

The Lagrangian in equation (\ref{chqmR}) contains couplings involving the
%mass matrix field  $\widetilde{M}_q$
the current mass term and the chiral quark fields. This
makes it possible to calculate the counter-terms  needed in
order to keep the chiral Lagrangian finite after the inclusion of chiral
loops. The counter-term for the factorizable part of the amplitude  has
been considered in \cite{ahjoe} when calculating $f_B$. In the case of
the non-factorizable part of the amplitude, we need to consider
similar diagrams as those shown in figure \ref{fig:bbargg}, with 
mass insertion like in figure \ref{fig:count}, where  mass insertion
is indicated by a cross on the light quark line.
 The bosonized current with mass insertion is
\bea
\left(\overline{q_L} t^a\,\gamma^\alpha Q_v^{(\pm)}\right)_{1G,m_q} 
\;   \longrightarrow \;  
\fr{G_H g_s}{32m\pi^2}\,\e^{\alpha\beta\mu\rho}(\pm v_\alpha) G_{\mu\rho}^a
\;Tr\left[\xi^\dagger \gamma^\alpha  L \, H_a^{(\pm)}
\left(\widetilde{M}_q^V\right)_{aq}\gamma_\beta\gamma_5\right] \; \, .
\label{curGM}
\eea
This result can also be obtained by simply differentiating the right hand 
side of equation (\ref{1G}) with respect to $m$.

The bosonized version of the $Q(\Delta B=2)$ operator can then be
split in a pseudo scalar and a vector part:
\bea
&&Q(\Delta B=2)_{\text{Bos.}} \,= A_P \,\B\,+\,A_V\,\Bs\,
\text{, where : }\nonumber\\
&&A_P=\fr{1}{2}(1+\fr{1}{N_c})(C_1-C_2)
\alpha_H^2\left(1+2\fr{\omega_1}{\alpha_H}m_q\right)-C_1\gc
\left(\beta_B+\omega_\beta m_q\right)\label{bosQ}\; ,\\ 
&&A_V=
\fr{1}{2}(1+\fr{1}{N_c})C_1\alpha_H^2\left(1+2\fr{\omega_1}{\alpha_H}
m_q\right) - \gc\left((C_1-\fr{C_2}{3})\beta_B
+C_1 \omega_\beta m_q\right)\nonumber \; .
\eea
 The quantity
$\omega_\beta$ is the counter-term obtained from (\ref{curGM}), and $\omega_1$
is a counter-term for $f_{B_s}$ found in \cite{ahjoe}:
\bea
&\omega_\beta&=\fr{G_H^2}{64\pi m}\left\{1+\fr{4\pi
f^2}{N_cm^2}\right\}\; ,\\
&\omega_1&=\fr{(1-3\ga)}{G_H}-\fr{(9\pi-16) G_H}{192 m^3}\gc \; .
\eea
For the current quark mass  entering (\ref{bosQ})
 we will use
\bea
 m_d=-m_\pi^2f^2/\qc \; \, , \; \text{and} \; \, m_s=-m_K^2f^2/\qc \; \, .
\label{curmas}
\eea
The term including the vector fields $P_\mu$ are needed in order to calculate 
chiral corrections where $B^*$ are included.
From equation the equations (\ref{matrQ}), (\ref{Bhat})  and (\ref{bosQ})
 the renormalization invariant 
bag parameter can be extracted. Anticipating the results of the two next 
sections, it can be written in the 
form:
\begin{equation}
\hat{B}_{B_q}=\fr{3}{4} \, \widetilde{b}
 \left[ 1 + \fr{1}{N_c} \left(1 - \delta_G^B(1+\fr{\tau_\chi^G}{32\pi^2f^2})
\right) + \fr{\tau_b}{m_b} + \left(1+\fr{1}{N_c}\right)
\fr{\tau_\chi}{32 \pi^2 f^2} \right] \; \, ,
\label{Bhatform}
\end{equation}
where 
\begin{equation}
\widetilde{b} \; = \; b(m_b) \, 
\left[ \fr{C_1-C_2}{(C_\gamma+C_v)^2}\right]_{\mu=\Lambda_\chi} \;  . 
\end{equation}
% At LO (neglecting $C_v$ and $C_2$) it turns out that  
%$\widetilde{b}(\mu)=b(m_b)$ because $C_1=(C_\gamma)^2$. 
%Thus, to LO,  $\widetilde{b}(\mu)$ is independent of
%the renormalization scale $\mu$ \cite{pol,vol}.
% At NLO this is not the case anymore.
We find from (\ref{bosQ}) the parameter due to genuine non-factorizable
 effects:
\begin{equation}
\delta_G^B \, = \,    N_c \gc \fr{\beta_B}{\alpha_H^2} \,
\left[ \fr{2 C_1}{C_1-C_2} \right]_{\mu=\Lambda_\chi}
 \; \, .
\end{equation}
Note that this parameter is formally of order $(N_c)^0$ and is positive,
which means that this non-factorizable contribution reduces the value 
of $\hat{B}$ according to (\ref{Bhatform}).
 Thus we are qualitatively in agreement  
with \cite{Mel}, where a negative contribution to the
 bag factor from gluon condensate effects  is found.

Using the 
relation between $\alpha_H$ and $f_B$ in Eq. (\ref{fb}) and the
expression value for $G_H$  in equation (\ref{gcgh}), we may also write :
\begin{equation}
\delta_G^B \, = \, 
 \frac{N_c \gc}{32\pi^2 f^2 f_B^2} \, \frac{m}{M_B} \, \rho \,
\left\{1+\fr{4\pi}{N_c}\left(\fr{f}{m}\right)^2
+\fr{8\pi^2}{N_c^2}\left(\fr{f}{m}\right)^4\right\} \,
\left[ \fr{ C_1}{C_1-C_2} \right]_{\mu=\Lambda_\chi} \; \ .
\end{equation}
Numerically, $f$ and $f_B$ are of the same order of magnitude, and 
$\delta_G^B$ is therefore suppressed like $m/M_B$ compared to the corresponding
 quantity 
\begin{equation}
\delta_G^K \, = \, N_c 
 \frac{\gc}{32\pi^2 f^4} \; ,
\end{equation}
for $K- \overline{K}$ mixing. However, one should note that $f_B$
scales as $1/\sqrt{M_B}$ within HQEFT, and therefore 
 $\delta_G^B$ is still formally of order  $(m_b)^0$.

The formula (\ref{Bhatform}) is a generalization of a similar formula
found for $K- \overline{K}$ mixing \cite{BEFL}.
 The quantities $\tau_b$ and $\tau_\chi^G$
 will be calculated in the next sections, while $\tau_\chi$ is known 
from previous work \cite{GrinWise}.
More specific, the quantity $\tau_b$, to be calculated in the next section,
 has dimension $(mass)^1$ and depend on
 hadronic parameters calculated within the HL$\chi$QM. Similarly, the quantity
 $\tau_\chi$ contains the chiral corrections to the bosonized versions of 
$Q_{1,2}$ to be presented in section VI. The quantity 
$\tau^G_\chi$ contains the chiral corrections proportional to
 $\gc$ and the counter-terms $\omega_\beta$ and $\omega_1$.

\section{$1/m_Q$ corrections}

The $1/m_Q$ corrections have been defined in equation
(\ref{O1}-\ref{S}). In the HL$\chi$QM we only need to consider
(\ref{O1}) and (\ref{S}). This is due to the fact that when we are
considering terms in the effective Lagrangian for $\bbar$ mixing 
the external particles carry no redundant momenta \cite{ahjoe}.
(In other words, the $B$-mesom momenta are $p_B=M_B v$). Hence the 
 operators in (\ref{O2}) and (\ref{O3}) will give  zero contribution. 

The operator in equation (\ref{O1}) can be written on the form
\begin{equation}\label{MatrO1}
X_1=2\sum_{j=1}^3\overline{q}\,\Gamma_j\, i\,D_\alpha\, \Qp\;
\overline{q}\,\Theta_j\, \Qm\,+\,
2\sum_{j=4}^6\overline{q}\,\Gamma_j\,  \Qp\;
\overline{q}\,\Theta_j\,\,i\,D_\alpha\, \Qm\, ,
\end{equation}
where $\Gamma^\alpha,\,\Theta$ are defined :
\bea
&\Gamma_1=R\,\gamma^\alpha  \,\,\,\,\qquad\qquad\qquad&\Theta_1=R
\nonumber\\
&\Gamma_2=R\,g^{\mu\alpha}
\,\,\qquad\qquad\qquad&\Theta_2=R\,\gamma_\mu
\nonumber\\
&\Gamma_3=-i\,\e^{\lambda\alpha\nu\rho}\,v^\lambda\,R\,\gamma^\nu 
  \qquad&\Theta_3=R\,\gamma^\rho
\nonumber\\
&\Gamma_4=R  \qquad\qquad\qquad\qquad&\Theta_4=R\,\gamma^\alpha
\nonumber\\
&\Gamma_5=R  \qquad\qquad\qquad\qquad&\Theta_5=R\, g^{\mu\alpha}
\nonumber\\
&\Gamma_6=-i\,\e^{\lambda\alpha\nu\rho}\,v^\lambda\,R\,\gamma^\rho 
  \qquad&\Theta_6=R\,\gamma^\nu\, ,	
\eea
where $D_\alpha$ is the covariant derivative containing the gluon field.
Note that the operator $X_1$ is Fierz symmetric \cite{mannel}.
We bosonize $X_1$ in the same way as $Q_{1,2}$.

Some two-quark operators appearing in (\ref{MatrO1}) are already studied 
in \cite{ahjoe}
when calculating $1/m_b$ corrections to $f_B$. We use those results 
when bosonizing $X_1$, and the result 
 can be written:
\begin{eqnarray}
&&X_1\to X_1^{\text{bos}}=
\nonumber\\
&&
 \sum_{i=1}^3\left\{2(1+  \fr{1}{N_c})\fr{\alpha_H}{2}
Tr\left[\xi^\dagger\Theta_i\,H_v^{(-)}\right]
\;\fr{1}{2}
Tr\left[\xi^\dagger\Gamma_i\,H_v^{(+)}\left(\alpha_3^\gamma\gamma^\alpha
+\alpha_3^vv^\alpha\right)\right] \nonumber\right.
\\&&\left.
+\, 4\beta_1  Tr\left[\xi^\dagger\Theta_i H_v^{(-)}
\left(-\beta_2\left\{\sigma^{\mu\nu},\gamma \cdot v \right\}+
\beta_4\sigma^{\mu\nu}\right)\right\}\;
 Tr\left[\xi^\dagger\Gamma_i H_v^{(+)}
\left(\beta_3D_{\mu\nu\alpha}+2m\beta_2\sigma_{\mu\nu}v_\alpha
\right)\right\}\right\}
\nonumber\\
&&
 +\sum_{i=4}^6\left\{2(1+  \fr{1}{N_c})\fr{\alpha_H}{2}
Tr\left[\xi^\dagger\Gamma_i\,H_v^{(-)}\left(\alpha_3^\gamma\gamma^\alpha
-\alpha_3^vv^\alpha\right)\right]\;
\fr{1}{2}
Tr\left[\xi^\dagger\Theta_i\,H_v^{(+)}\right] \label{bosO1}
\right.\\&&\left.
+\, 4 \beta_1Tr\left[\xi^\dagger\Gamma_i H_v^{(-)}
\left(\beta_3D_{\mu\nu\alpha}-2m\beta_2\sigma_{\mu\nu}v_\alpha
\right)\right]\;
 Tr\left[\xi^\dagger\Theta_i H_v^{(+)}
\left(\beta_2\left\{\sigma^{\mu\nu},\gamma \cdot v \right\}+
\beta_4\sigma^{\mu\nu}\right)\right]\ket\right\} \; \, , \nonumber
\end{eqnarray}
where $D_{\mu\nu\alpha} \eq \left\{\sigma_{\mu\nu},\gamma_\beta\right\}
\left(g_{\alpha\beta}-v_{\alpha}v_{\beta}\right)$.
The second and fourth lines  are genuinely non-factorizable.
The $\alpha$'s and $\beta$'s are hadronic parameters calculated within the
HL$\chi$QM, and are given in Appendix \ref{param}.
Evaluating the sums and traces in equation (\ref{bosO1}) we arrive
at :
\bea
X_1^{\text{bos}}=\left\{\alpha_H \alpha_3^\gamma(1+\fr{1}{N_c})
 + \gc\beta_B^{(2)}\right\}
\left(-\Bs+3\B\right)\; ,
\eea
where $\beta_B^{(2)}$ is a combination of the $\beta_i$'s and can be written
\bea
\beta_B^{(2)}\eq\fr{\pi}{4N_c} (1-\ga) \left(1+\fr{4\pi}{3N_c}\fm^2\right)
\; \, .
\eea

The bosonating of the nonlocal operators is rather straight forward
in this model. The result for the factorizable part of the non local
operators can be found in \cite{ahjoe} in the calculation of $f_B$ :
\bea
\sum_{i=1}^4\fr{A_iS_i^{\text{Fact}}}{m_b}
\to&&
-\left(1+\fr{1}{N_c}\right)\fr{\alpha_H}{m_b G_H}
\left(\mu_\pi^2-d_{\cal M}\fr{\mu_G^2}{3}\right)
\nonumber \\&&
 \times \, \left[C_1\Bs
+(C_1-C_2)\B\right] \; \, .
\eea
The result for the nonfactorizable part of the operators is :
\bea
&&\sum_{i=1}^4\fr{A_iS_i^{\text{Nfact}}}{m_b}
\to\nonumber \\
&&\fr{1}{m_b}\gc\beta_K\left((C_1-\fr{1}{3}C_2)\Bs+C_1\B\right)\nonumber\\
+&&
\fr{1}{m_b}\gc C_M \left(C_1  \beta_M^{(1)}+C_2  \beta_M^{(2)}\right)
%\nonumber \\&&
\left[-\Bs
+3\B\right]\; ,\label{beta_K}
\eea
where the quantities $\beta_{K}$ and $\beta_{M}^{(1,2)}$'s are given
 in Appendix \ref{param}.

We need 
$f_B$ which has been calculated in \cite{ahjoe} to $1/m_b$ :
\bea
&&f_H\sqrt{M_H}=\alpha_H(C_\gamma+C_v)\left(1+\fr{\kappa_b}{m_b}
+\fr{\kappa_\chi}{32\pi^2f^2}\right) \; \, , \; \;
  \text{where :}\,\nonumber \\
&&\kappa_b=-\fr{(\e_1-6\e_2)}{2}+
\fr{(B_\gamma\alpha_3^\gamma+B_v\alpha_3^v)}{2\alpha_H (C_\gamma-C_v)}
-\fr{(\mu_\pi^2-\mu_G^2)}{G_H \alpha_H}\nonumber\\
&&\kappa_{\chi_d}=-\fr{11}{18}\left\{
-m_K^2(1+\ga^2) 
+m_K^2(\ln\fr{m_K^2}{\mu^2}+\fr{2}{11}\ln\fr{4}{3})(1+3\ga^2) \right\} \; , \\
&&\kappa_{\chi_s}=-\fr{13}{9}\left\{-m_K^2(1+\ga^2)
+m_K^2(\ln\fr{m_K^2}{\mu^2} + \fr{4}{13}\ln\fr{4}{3})(1+3\ga^2)\right\}
+\fr{\omega_132\pi^2f^2}{\alpha_H}m_s \; , \nonumber
\eea
where $B_\gamma$ and $B_v$ are sums of Wilson coefficients. The
contribution to the  bag
parameter from $1/m_b$ corrections  
can now be extracted (see eq. (\ref{Bhatform})):
\bea
\tau_b = &&\left(1+\fr{1}{N_c}\right)
\left\{\fr{\alpha_3^\gamma}{\alpha_H}\left(\fr{6B_1}{C_1-C_2}
-\fr{B_\gamma}{C_\gamma+C_v}
\right)-\fr{\alpha_3^v}{\alpha_H}\fr{B_v}{(C_\gamma+C_v)}\right\}\nonumber\\
&&+\fr{6C_1}{(C_1-C_2)\alpha_H^2}\gc\left\{\fr{B_1}{C_1}\beta_B^{(2)}
+\fr{\beta_K}{3}+C_M\beta_M^{(1)}+\fr{C_2C_M}{C_1}\beta_M^{(2)}\right\}
\; \, .
\eea
 It should be noted that $1/m_b$
corrections increases  $\hat{B}$, in agreement with \cite{mannel}.

\section{Chiral corrections}

\begin{figure}[t]
\begin{center}
   \epsfig{file=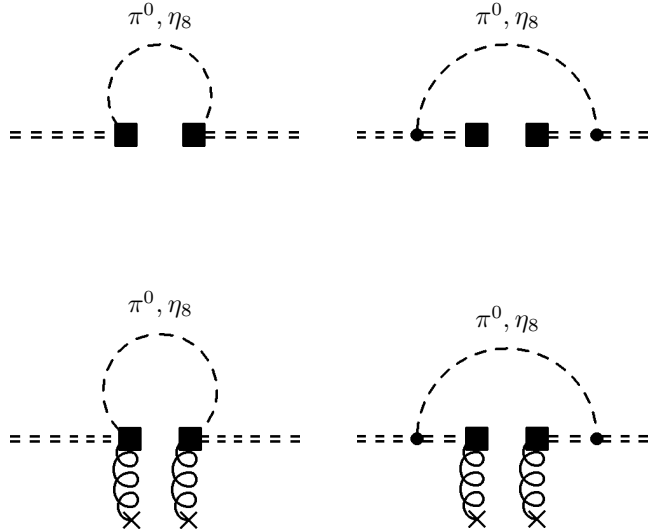}
\caption{Diagrams contributing to the bag parameter}
\label{fig:bagparam}
\end{center}
\end{figure}

We will only consider chiral corrections to $Q_{1,2}$ in equation
(\ref{Q1}) and (\ref{Q2}). Adding chiral corrections to operators
proportional to $1/m_Q$ will be considered as  higher order. The
chiral corrections to the bag parameter have been
considered in \cite{GrinWise}. Some of the corrections are simply 
corrections to $f_{B_q}$ \cite{GriBo,goity,cheng2}. 
The diagrams shown in
figure \ref{fig:bagparam} are those which are genuinely non-factorizable, 
i.e. they are not included in chiral corrections to $f_{B_q}$.

\begin{figure}[t]
\begin{center}
   \epsfig{file=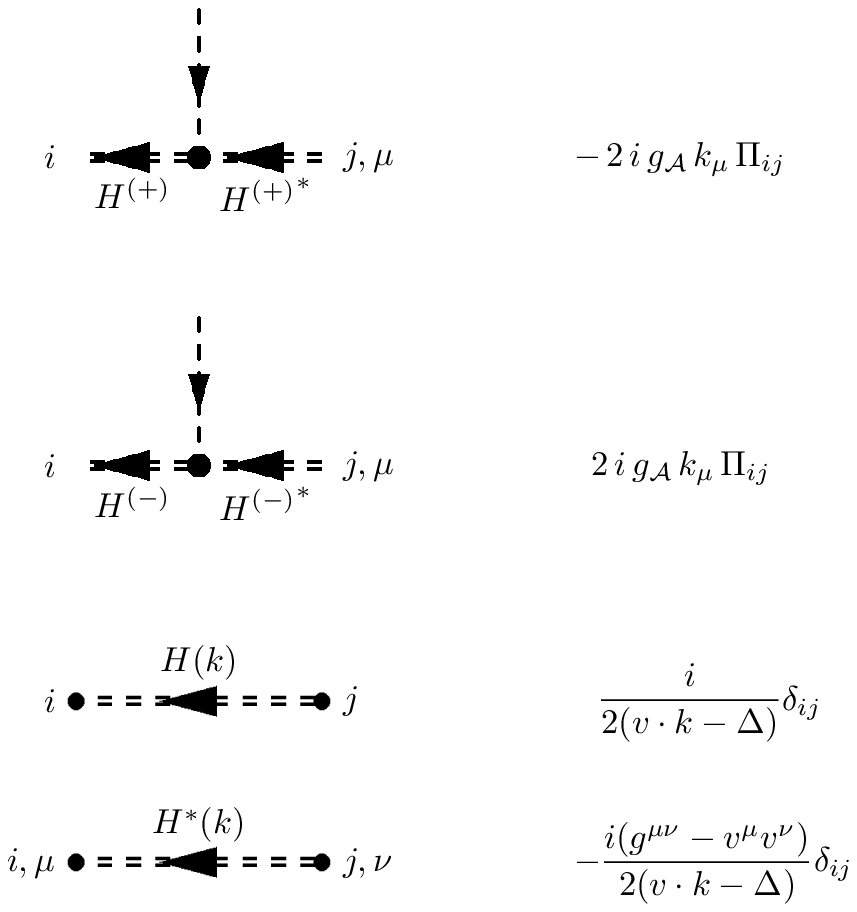}
\caption{Feynman rules for the strong sector, $\Pi$ is given in
equation (\ref{Pi})}
\label{fig:loop5}
\end{center}
\end{figure}

The chiral corrections ($\tau_\chi$) to the bag parameter can then be
 written :
\bea
\tau_\chi=&&d_\chi\left\{-\fr{2}{9}m_K^2\ln\left(\fr{4m_K^2}{3\mu^2}\right)
-\fr{2}{9}m_K^2
\right.\\&&\left.+
\fr{C_1}{C_1-C_2}\ga^2\left((\fr{2}{3}m_K^2
-\Delta^2)\ln\left(\fr{4m_K^2}{3\mu^2}\right)
-\fr{8}{9}m_K^2+\fr{8}{3}\Delta^2(\,2-3F(\Delta/m_\eta)\,)\right)\right\}
\; \, ,\nonumber\\
\tau^G_\chi=
&&d_\chi\left\{-\fr{2}{9}m_K^2\ln\left(\fr{4m_K^2}{3\mu^2}\right)
-\fr{2}{9}m_K^2
\nonumber\right.\\&&\left.+
\fr{C_1-C_2/3}{C_1}\ga^2\left((\fr{2}{3}m_K^2
-\Delta^2)\ln\left(\fr{4m_K^2}{3\mu^2}\right)
-\fr{8}{9}m_K^2+\fr{8}{3}\Delta^2(\,2
-3F(\Delta/m_\eta)\,)\right)\right\}\nonumber\\
&&- d_s\left(\fr{\omega_\beta}{\beta_B}+
2\fr{\omega_1}{\alpha_B}\right)\, 32 \pi^2 f^2 \, m_s \; \, ,
\eea
where we have ignored the pion mass  and used the mass relations
$m_{\eta_8}^2=4m_K^2/3$. The function
$F(x)$ is defined in equation (\ref{F}) and :
\begin{equation}
d_\chi=\begin{cases} 1\quad\text{for}\quad B_d\\
                     4\quad\text{for}\quad B_s\end{cases}
\quad\text{and}\quad
d_s=\begin{cases} 0\quad\text{for}\quad B_d\\
                     1\quad\text{for}\quad B_s\end{cases}
\end{equation} 
If one ignores the counter-term given by 
 $\omega_\beta$, and take the limit
 $\Delta \equiv  M_H^* -M_H \to 0$, 
we obtain the same result as in \cite{GrinWise}. For 
the bare coupling constant $f$ we will use the value $f$=86 MeV \cite{cheng2}.
The Feynman rules for chiral loops are given in figure  \ref{fig:loop5}.

\section{Numerical Results}

\begin{figure}[t]
\begin{center}
   \epsfig{file=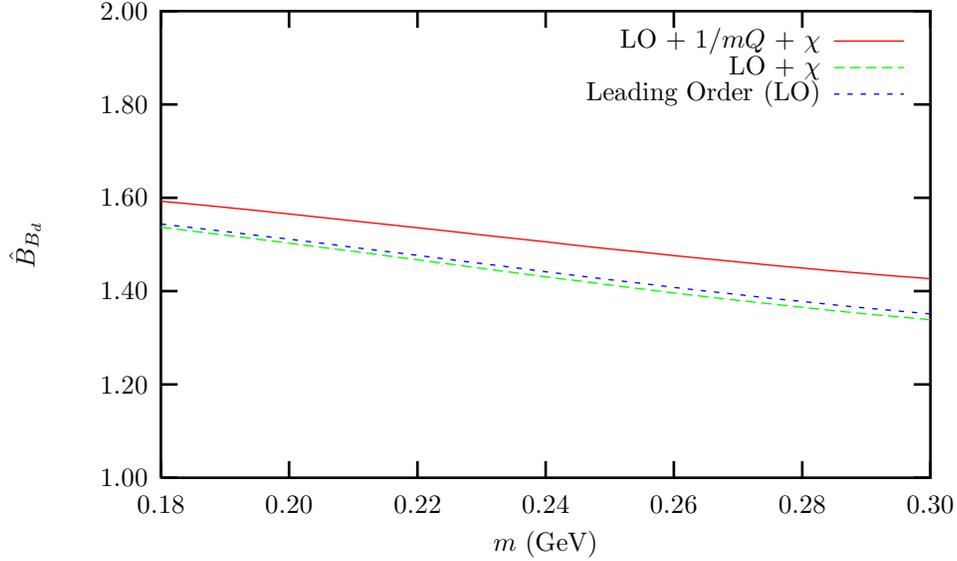}
\caption{The bag parameter $\hat{B}$ for $B_d$}
\label{fig:Bb}
\end{center}
\end{figure}

\begin{figure}[t]
\begin{center}
   \epsfig{file=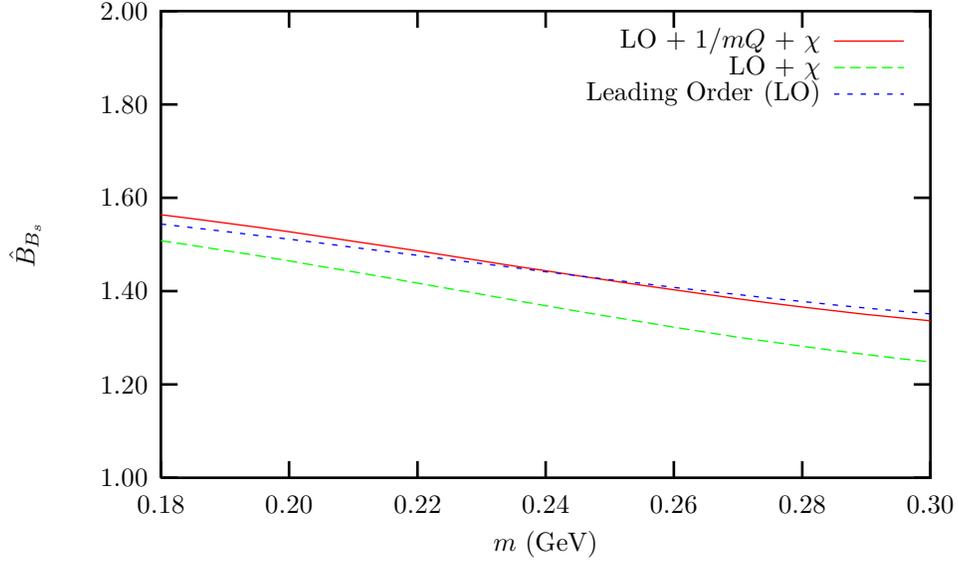}
\caption{The bag parameter $\hat{B}$ for $B_s$}
\label{fig:Bbs}
\end{center}
\end{figure}

The model dependent parameters of the HL$\chi$QM was fixed in \cite{ahjoe}
by using various constraints. For instant, the
 constituent light quark mass  was determined
 to be 
$m=220\pm 30\,~$MeV.
Using the parameters from \cite{ahjoe}, we obtain
%is what we need to calculate the bagparameter. The result 
(using $\Delta=M_H^*-M_H=0.025~\text{GeV}$):
\bea
&\tau_b=(0.26\pm0.04)\, \text{GeV} \qquad
&\delta_G^B=(0.5\pm0.1)\nonumber\\
&\tau_{\chi_d}=-(0.02\pm0.01)\, \text{GeV}^2 \qquad
&\tau_{\chi_s}=-(0.10\pm0.04)\, \text{GeV}^2\nonumber\\
&\tau^G_{\chi_d}=-(0.03\pm0.01)\, \text{GeV}^2 \qquad
&\tau^G_{\chi_s}=(0.12\pm0.06)\, \text{GeV}^2\nonumber\\
&\hat{B}_{B_d}=1.53\pm0.05 \qquad
&\hat{B}_{B_s}=1.48\pm0.08
\nonumber\\
&f_{B_d}=(170\pm 25)\,\text{MeV} \qquad &f_{B_s}=(180\pm
25)\,\text{MeV}\nonumber\\
&f_{B_d}\sqrt{\hat{B}_{B_d}}=(215\pm 30)\,\text{MeV} \qquad
&f_{B_s}\sqrt{\hat{B}_{B_s}}=(225\pm 30)\,\text{MeV}\nonumber\\
&\xi=\fr{f_{B_s}\sqrt{\hat{B}_{B_s}}}{f_{B_d}\sqrt{\hat{B}_{B_d}}}
= 1.05\pm 0.01\qquad  &
\fr{f_{B_s}}{f_{B_d}}=1.08\pm0.02
\label{Res1}
\eea
The decay constants $f_{B_d}$ and $f_{B_s}$ were also given 
in \cite{ahjoe}, but are listed also 
here for completeness. (Note, however, that the values are slightly different,
because in   \cite{ahjoe} we did not distinguish $f_\pi$ from the bare
coupling $f$.)
The values for the bag parameter $\hat{B}$ are in agreement
 with  lattice
calculations \cite{latt,Mar-latt}.
A plot of $\hat{B}$  as a function of the constituent quark
mass $m$ is shown in figure \ref{fig:Bb} and  \ref{fig:Bbs}.
We observe that the values of $\hat{B}$ are
 fairly stable over a
large variation of light quark constituent mass $m$. Especially this is
the case for $B_d$.
 From $m=180$ MeV and $m=300$ MeV the
bag factors only changes with $10\%$.
We note that $1/m_b$ corrections are small.

The values for the $f_B$'s and especially for the ratio $f_{B_s}/f_{B_d}$
(and $\xi$) in (\ref{Res1})
are  a bit  low \cite{latt,Ljublj}.
There might be at least three reasons for this.
First, concerning the absolute value for $f_B$'s, they
 dependent significantly on the
value of the quark condensate, as seen from  equation
 (\ref{qcrel}) and (\ref{fb}). In
\cite{ahjoe} we used the ``standard'' value $\qc=(-240\,\text{MeV})^3$,
without any uncertainty. 
 It could be argued that
we should have used an uncertainty of 10 MeV, say,
 for $\qc^{1/3}$, although the
wide range 190 to 250 MeV used for
$m$ will to some extent compensate for this.
Second, it might be that our expansion within the HL$\chi$QM overestimates the
counter-term $\omega_1$ which reduces  $f_{B_s}$.
However, neglecting this counter-term would give
the high value $f_{B_s}/f_{B_d} \simeq 1.3$. 
Third, our value for the axial pion coupling $\ga$ in  
(\ref{LS1}) might be too low. 
%Unfortunately, there is not enough precise
%experimental data to constrain the parameters,
In 
\cite{ahjoe} we used input from QCD sum rules \cite{beyalev} both in the
$B$- and $D$-sectors.
Alternatively, we may use the experimental value for the effective
coupling $\ga^{H^*H\pi}=0.59 \pm 0.09$ in the $D$-sector \cite{anas},
giving almost the same bare coupling $\ga=0.59 \pm 0.04$. Using 
this bare coupling also in the $B$-sector (instead of $\ga=0.42 \pm 0.06$
in \cite{ahjoe}), and in addition  $\qc^{1/3}= 
(-240 \pm 10) \,\text{MeV}$,
 we obtain an alternative set of values:
\bea
&\tau_b=(0.25\pm0.04)\, \text{GeV} \qquad
&\delta_G=(0.5\pm0.2)\nonumber\\
&\tau_{\chi_d}=-(0.06\pm0.01)\, \text{GeV}^2 \qquad
&\tau_{\chi_s}=-(0.25\pm0.04)\, \text{GeV}^2\nonumber\\
&\tau^G_{\chi_d}=-(0.07\pm0.01)\, \text{GeV}^2 \qquad
&\tau^G_{\chi_s}=(0.2\pm0.2)\, \text{GeV}^2\nonumber\\
&\hat{B}_{B_d}=1.51\pm0.09 \qquad
&\hat{B}_{B_s}=1.37 \pm 0.14
\nonumber\\
&f_{B_d}=(190\pm 50)\,\text{MeV} \qquad &f_{B_s}=(210\pm
70)\,\text{MeV}\nonumber\\
&f_B\sqrt{\hat{B}_{B_d}}=(240\pm 70)\,\text{MeV} \qquad
&f_{B_s}\sqrt{\hat{B}_{B_s}}=(260 \pm 90)\,\text{MeV}\nonumber\\
&\xi=\fr{f_{B_s}\sqrt{\hat{B}_{B_s}}}{f_{B_d}\sqrt{\hat{B}_{B_d}}}
= 1.08\pm 0.07  \qquad &
\fr{f_{B_s}}{f_{B_d}}=1.14\pm0.07
\label{Res2}
\eea 
We observe that the value for $f_{B_s}/f_{B_d}$
in (\ref{Res2}) is close to the standard one.
%For the values in (\ref{Res2}), the analog  curves of those in  figures 
%\ref{fig:Bb} and \ref{fig:Bbs} would look very much the same.

%The decay constants increases compared to (\ref{Res1}) , but the bag
% factors are quite stable.
%They only increase with roughly $5\%$. 
%As the result (\ref{Res2}) includes (\ref{Res1}) by taking into account 
%the error bars, we take (\ref{Res2})
%as our final result.

To conclude, we have calculated the bag parameter $\hat{B}$ for the $B_d$ and 
$B_s$ mesons. Combining our two alternative sets of values
(and consider the range of values) we find
$\hat{B}_{B_d} = 1.51 \pm 0.09$ and  $\hat{B}_{B_s} =1.40 \pm 0.16$.
 The value for
 $\hat{B}_{B_s}$ is more sensitive to chiral loops and counter-terms,
and therefore the uncertainty is bigger.

In principle, $\hat{B}$ is renormalization invariant
 ($\mu$ independent). This cannot be shown within our approach. By 
construction,
 perturbative QCD within HQEFT, the HL$\chi$QM and chiral perturbation 
theory are  matched at the scale $\Lambda_\chi$. However,
 we have a reasonable
good matching numerically as in \cite{BEF}.
Varying the renormalization scale
$\mu = \Lambda_\chi$ in the range 0.8 GeV to 1 GeV,   the bag
parameters only change with 6$\%$. 
Moreover, like in \cite{BEFL}, the formula (\ref{Bhatform})
 nicely shows  the the various parts
building up the total result for $\hat{B}$.

\appendix

\section{Loop integrals}\label{app:loop}

The divergent integrals entering in the bosonization of the 
HL$\chi$QM are defined :
\bea
I_1 \, &\eq &\,\int\fr{d^dk}{(2\pi)^d}\fr{1}{k^2 -  m^2} \\
I_{3/2}\, &\eq &\, \int\fr{d^dk}{(2\pi)^d}\fr{1}{(v\cdot k)(k^2 -  m^2)} \\
I_2\, &\eq\, &\int\fr{d^dk}{(2\pi)^d}\fr{1}{(k^2 -  m^2)^2} 
\eea
The integrals needed in the calculation of chiral corrections to the
bag parameter are :
\bea
L^{m,\Delta}_{1,1}&&
=\int\,\fr{d^dk}{(2\pi)^d}
\fr{1}{(k^2 - m^2)(v\cdot k-\Delta)}
%\nonumber \\&&
=\fr{-i \Delta}{8\pi^2}\left(
\fr{1}{\ebar}-\ln(m^2)+2-2F(m/\Delta)\right)\label{L11}
\eea
\bea
&&\int\,\fr{d^dk}{(2\pi)^d}
\fr{k^{\mu}k^{\nu}}{(k^2 - m^2)(v\cdot k-\Delta)}=
A\,g^{\mu\nu}+B\,v^{\mu}v^{\nu}\nonumber \\
&&A=\fr{1}{d-1}\int\,\fr{d^dk}{(2\pi)^d}
\fr{k^2-(v\cdot k)^2}{(k^2 - m^2)(v\cdot k-\Delta)}\nonumber\\
&&=\fr{i\Delta}{16\pi^2}\left\{(-\fr{1}{\ebar}+
\ln(m^2)-1)(m^2-\fr{2}{3}\Delta)-\fr{4}{3}F(m/\Delta)(\Delta^2-m^2)
-\fr{4}{3}(m^2-\fr{5}{6}\Delta^2)\right\}\\
&&B=-A+\int\,\fr{d^dk}{(2\pi)^d}
\fr{(v\cdot k)^2}{(k^2 - m^2)(v\cdot k-\Delta)}\nonumber\\
&&=\fr{-i\Delta}{16\pi^2}\left\{(-\fr{1}{\ebar}+
\ln(m^2)-1)(2m^2-\fr{8}{3}\Delta)-\fr{4}{3}F(m/\Delta)(4\Delta^2-m^2)
\right.\nonumber\\ &&\left.\qquad\qquad
-\fr{4}{3}(m^2-\fr{7}{3}\Delta^2)\right\}\label{Lmunu}
\eea
where:
\begin{equation}\label{F}
F(x)=\begin{cases}
&-\sqrt{x^2-1}\tan^{-1}(\sqrt{x^2-1})\qquad x>1\\
&\,\sqrt{1-x^2}\tanh^{-1}(\sqrt{1-x^2})\qquad x<1
\end{cases}
\end{equation}
In the case of $\Delta>m$ we have ignored an analytic 
real part in (\ref{L11}). Equation (\ref{L11}) coincides with the one
obtained in \cite{GriBo} however equation (\ref{Lmunu}) differs
by a factor $-2/3(m^2-2/3\Delta^2)$ inside the parenthesis of the
expressions for $A$ and $B$. This is presumably due to the factor
$1/(d-1)=(1-2/3\e)/3$ in $A$.  

\section{Some detailed expressions for hadronic parameters}\label{param}

The parameters of equation (\ref{bosO1}) are :
\bea
&\alpha_3^\gamma&\eq\fr{m}{3}\alpha_H+\fr{G_H}{6}\qc\nonumber\\
&\alpha_3^v&\eq\fr{m}{3}\alpha_H+\fr{2}{3}G_H\qc\nonumber\\
&\beta_1&\eq\fr{G_B^2\pi^2}{12}\gc\nonumber\\
&\beta_2&\eq-\fr{f^2}{4m^2N_c}\nonumber\\
&\beta_3&\eq-\fr{\delta\ga}{4G_B^2N_c}\nonumber\\
&\beta_4&\eq\fr{1}{8\pi}\label{mqparam}
\eea

The $\beta_{K,M}^{(1,2)}$'s in (\ref{beta_K})
are given by :

\bea
\beta_K^{(1)}&=&\fr{m}{256\pi^2}G_B^2\left\{1+\fr{4}{\pi}-\fr{8\pi}{N_c}\fm^2
(1+\fr{1}{\rho}-\pi)-\fr{32\pi^2}{N_c^2}\fm^4\right.\nonumber \\
&&\left.
\qquad\qquad\quad-C_K\left[\fr{8\pi}{N_c}\fm^2+\fr{16\pi^2}{N_c^2}\fm^4\right]
\;\right\}\\
\beta_M^{(1)}&=&-\fr{\pi^2}{12N_c^2}\left(\fr{f}{m}\right)^2\\
\beta_M^{(2)}&=&\fr{\pi}{24N_c}\left\{1+\fr{2\pi}{N_c}\left(\fr{f}{m}\right)^2\right\}
\eea

%\section{Feynman rules}

\bibliographystyle{unsrt}

\end{document}